# Direct Imaging and Gradient-Based Analysis of the 12 August 2026 Partial Solar Eclipse from a Freely Rotating High-Altitude Balloon

Björn Poppe, Enno Gronewold, Matti Gehlen, Jona Schrader, Delia Gauk, Lisa Cordes, Maike I. Schmitz, Peter Schönfeld, Simon Jäger and Gerhard Drolshagen

Division for Radiation Physics and Space Environment, University Observatory, Carl von Ossietzky Universität Oldenburg, Germany

## Abstract

We report a proof-of-concept observation of the partial solar eclipse of 12 August 2026 using a freely rotating high-altitude balloon launched from Oldenburg, northern Germany. The payload carried two Insta360 ONE RS cameras equipped with 4K Boost wide-angle lenses; covered by filter material taken from a BRESSER eclipse viewing glass. Interval photographs were acquired every 10 s at ISO 800 and 1/1000 s. Of approximately 1,300 images, 21 contained a directly visible image of the eclipsed Sun, spanning 19:19:15-20:38:21 CEST and both sides of the local eclipse maximum. To test whether quantitative eclipse information could be recovered from these small, non-stabilized wide-angle images, the visible solar area was estimated using a two-dimensional gradient-based solar edge analysis and normalized to the first observation. The image-derived obscuration followed the independently calculated eclipse geometry with Pearson r = 0.959, a mean absolute difference of 9.5 percentage points, and an RMSE of 11.5 percentage points. Systematic deviations are consistent with the limitations of an uncalibrated action-camera system, including point-spread function, field-dependent lens response, filter geometry, and camera-to-camera differences. The results demonstrate that passive payload rotation can yield both visually useful and semi-quantitative direct eclipse observations without active solar pointing, and define a calibration strategy for future balloon-borne eclipse measurements.

## 1. Introduction

High-altitude balloons provide a comparatively inexpensive platform for atmospheric and astronomical observations. Solar eclipses are particularly attractive targets because a balloon payload can potentially record both the eclipsed Sun and large-scale atmospheric or terrestrial effects associated with the lunar shadow.

Balloon-borne eclipse observations were performed extensively during other solar eclipses. During the eclipse of 21 August 2017. Madhani et al. [1] used a high-altitude balloon and ground-based photodiode arrays to investigate eclipse shadow bands. The campaign also obtained solar-filtered camera observations of the partial phases. Direct solar imaging from a freely suspended payload is challenging because balloon gondolas can rotate rapidly. Peters [2] therefore described a dedicated solar-eclipse video payload with automatic solar pointing, while Bowman et al. [3] developed an active camera-stabilization system for maintaining a solar bearing from a moving high-altitude balloon.

The total solar eclipse of 12 August 2026 stimulated several new balloon campaigns. In Spain, coordinated projects including AstroCuenca launched camera-equipped sondes to image the eclipse and the lunar shadow from high altitude [4]. First imagery and mission reports have already appeared, while detailed scientific analyses of these datasets are still forthcoming. The present experiment provides a complementary observation from Oldenburg, Germany, outside the path of totality, where the eclipse reached approximately 86.3% obscuration.

## 2. Materials and Methods

The payload was prepared at Freie Schule Oldenburg, Oldenburg, Germany (approximately 53.126° N, 8.231° E). Image acquisition started immediately before release; the balloon was launched at approximately 19:20 CEST on 12 August 2026. A Stratoflights Weather Balloon 1600 (nominal balloon mass 1600 g) was filled with approximately 2400 liters of helium. The manufacturer specifies an average burst altitude of 36 km and an average burst diameter of approximately 11.1 m for this balloon type [5]. The flight reached a burst altitude of approximately 35 km.

The instruments were housed in a Stratoflights expanded-polystyrene probe box (external dimensions 22 × 22 × 18 cm; approximately 200 g) [6]. Recovery was supported by the Stratoflights GPS/mobile-network tracking system together with an additional GPS tracking device.

The imaging payload consisted of two Insta360 ONE RS cameras equipped with the 4K Boost Lens. The module provides a wide-angle view with a nominal 16-mm full-frame-equivalent focal length and f/2.4 aperture. Both cameras were operated in interval-photography mode with fixed settings of ISO 800 and 1/1000 s and an acquisition interval of 10 s.

The viewing direction was covered by filter material taken from one side of a BRESSER solar-eclipse viewing glass. The BRESSER eclipse glasses are specified for direct solar observation and comply with EN ISO 12312-2:2015 [7]. The camera payload had no active solar pointing or mechanical stabilization.

Approximately 1,300 photographs were acquired. After recovery, the complete image set was screened for the compact orange-red solar image produced by the filter. Twenty-one frames containing the Sun were identified. For quantitative analysis, 200 × 200 pixel crops centred on the solar image were extracted from the original photographs. No sharpening, local contrast enhancement, interpolation or synthetic reconstruction was applied to the images used for measurement.

Because the payload orientation changed continuously, the solar image appeared at different positions. Acquisition times were read directly from the EXIF metadata. For presentation in Figure 1, the crops were translationally registered to a common solar position; the quantitative area analysis was performed on the individual original crops.

For each timestamp, the geometrical fraction of the apparent solar disk occulted by the Moon was calculated topocentrically for the launch location from the apparent Sun-Moon separation and angular radii. The calculation gives a local maximum obscuration of 86.32% at approximately 20:08:53 CEST.

To test the potential of the acquired images for further scientific analysis, an independent image-based estimate was obtained using a gradient-based solar edge analysis. Each crop was converted to grayscale and mildly Gaussian-smoothed to suppress pixel-scale noise. The two-dimensional intensity-gradient magnitude was calculated from the horizontal and vertical image derivatives as

$$G(x,y) = \sqrt{(\frac{\partial I}{\partial x})^2 + (\frac{\partial I}{\partial y})^2}.$$

The apparent solar edge was associated with the local maximum of $G(x,y)$, corresponding approximately to the region of the strongest intensity transition between the solar image and its surroundings. The characteristic intensity of this maximum-gradient region was then used to delineate the largest connected solar region and determine its visible area $A_i$ in pixels. The first frame at 19:19:15 CEST was used as the reference area $A_1$ and was assigned its independently calculated geometrical obscuration $O_1^{\mathrm{geo}}$ = 2.71%. The image-derived obscuration for frame i was calculated as

$$O_i^{\mathrm{img}}[\%] = 100 - (100 - O_1^{\mathrm{geo}}[\%]) \cdot \frac{A_i}{A_1}$$

This normalization avoids requiring an independently calibrated unocculted solar area, which we did not find in our data. No frame-specific tuning to the expected eclipse geometry and no photometric, atmospheric-extinction, vignetting or point-spread-function correction was applied. The analysis was deliberately kept simple as a proof of concept for extracting quantitative eclipse information from non-stabilized wide-angle balloon images.

## 3. Results

Twenty-one direct solar images were identified among approximately 1,300 interval photographs (about 1.6%). The sequence extends from 19:19:15 CEST, when the calculated solar obscuration was 2.71%, to 20:38:21 CEST, when the obscuration had decreased to 31.98%, and therefore documents both the increasing and decreasing partial phases. Figure 1 shows the complete sequence. The closest recorded frame before maximum was acquired at 20:05:07 CEST and corresponds to a calculated obscuration of 83.95%.

The gradient-based solar edge analysis reproduced the overall eclipse progression. Across all 21 frames, image-derived and geometrically calculated obscuration were strongly correlated (Pearson r = 0.959). The mean absolute difference was 9.5 percentage points and the root-mean-square difference was 11.5 percentage points. Agreement was particularly close in several frames, including 19:25:13 (9.59% geometrical versus 8.91% image-derived), 19:29:14 (15.49% versus 16.48%), 19:35:15 (25.70% versus 21.01%), and the near-maximum frame at 20:05:07 (83.95% versus 80.03%). Systematic underestimation was more pronounced in parts of the later sequence. The complete numerical comparison is given in Table 1.

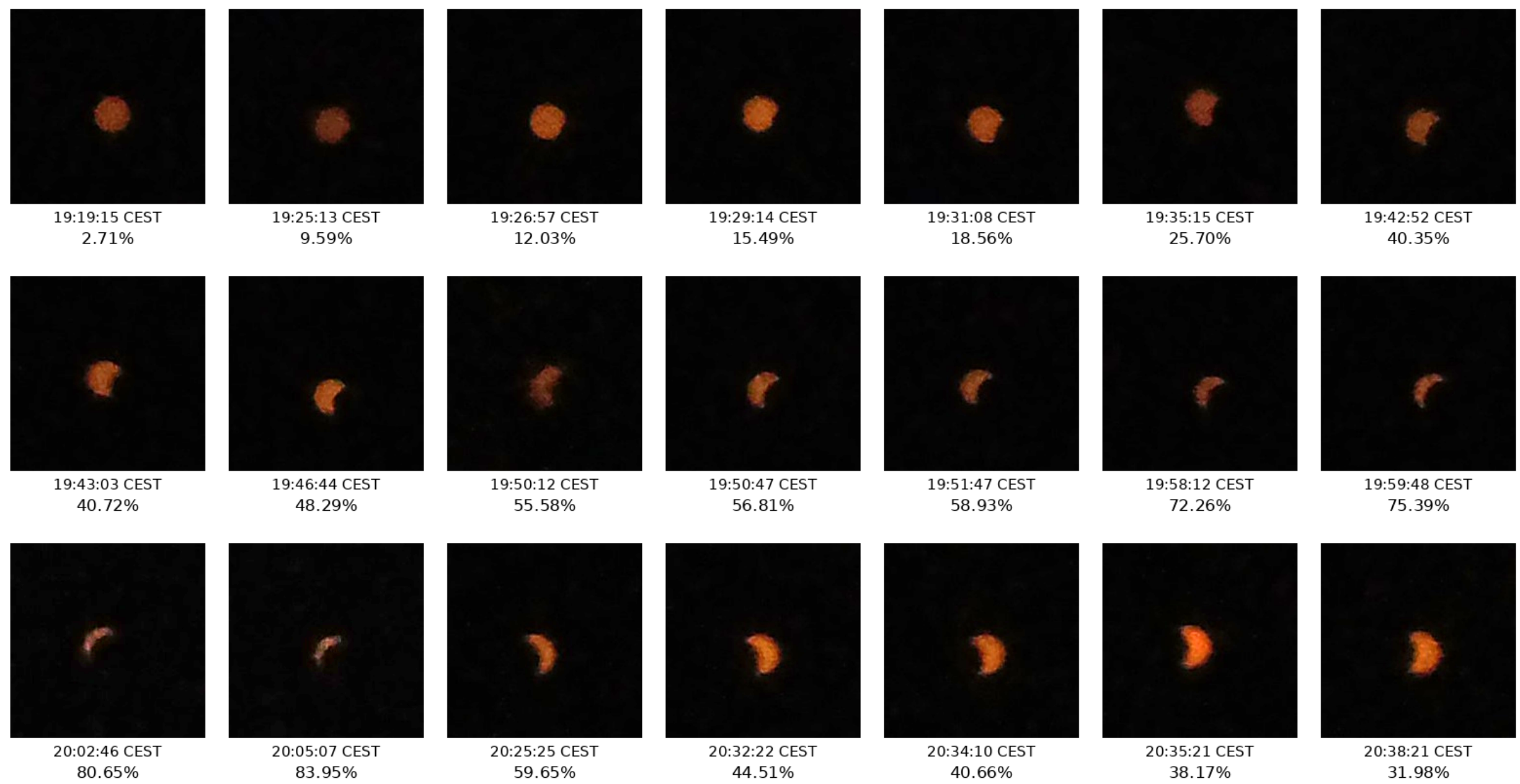

Figure 1. Sequence of 21 direct images of the partial solar eclipse of 12 August 2026 obtained during a high-altitude balloon flight launched from Oldenburg, Germany. Panels are the original 200 × 200 pixel crops on their black image background. For print presentation only, brightness was increased; the two particularly faint near-maximum frames at 20:02:46 and 20:05:07 CEST received a stronger brightness adjustment. No sharpening, geometric reconstruction, interpolation or alteration of the eclipse geometry was performed. Labels show the original EXIF acquisition time (CEST) and independently calculated geometrical solar obscuration. Image credit: AG Medizinische Strahlenphysik, University of Oldenburg.

Table 1. Geometrically calculated and image-derived solar obscuration using the gradient-based solar edge analysis.

| Time (CEST) | Geometrical obscuration (%) | Image-derived obscuration (%) | Difference (pp) |
|---|---|---|---|
| 19:19:15 | 2.71 | 2.71 | +0.00 |
| 19:25:13 | 9.59 | 8.91 | -0.68 |
| 19:26:57 | 12.03 | 7.33 | -4.70 |
| 19:29:14 | 15.49 | 16.48 | +0.99 |
| 19:31:08 | 18.56 | 8.02 | -10.53 |
| 19:35:15 | 25.70 | 21.01 | -4.69 |
| 19:42:52 | 40.35 | 35.67 | -4.69 |
| 19:43:03 | 40.72 | 24.84 | -15.88 |
| 19:46:44 | 48.29 | 42.16 | -6.14 |
| 19:50:12 | 55.58 | 40.78 | -14.80 |
| 19:50:47 | 56.81 | 52.09 | -4.72 |
| 19:51:47 | 58.93 | 52.19 | -6.74 |
| 19:58:12 | 72.26 | 57.31 | -14.96 |
| 19:59:48 | 75.39 | 65.77 | -9.62 |
| 20:02:46 | 80.65 | 72.85 | -7.80 |
| 20:05:07 | 83.95 | 80.03 | -3.92 |
| 20:25:25 | 59.65 | 40.49 | -19.17 |
| 20:32:22 | 44.51 | 22.58 | -21.92 |
| 20:34:10 | 40.66 | 24.84 | -15.82 |
| 20:35:21 | 38.17 | 26.52 | -11.65 |
| 20:38:21 | 31.98 | 12.25 | -19.73 |

## 4. Discussion

The observations demonstrate a deliberately simple alternative to active solar tracking. Rather than compensating for payload rotation, the experiment used a wide field of view and frequent interval photography. Natural rotation and oscillation effectively scanned the sky, and occasional intersections between the filtered camera field and the solar direction produced usable eclipse frames. This approach differs from dedicated pointing systems developed for previous eclipse balloon missions [2,3], but it has a precedent in the 2017 Pittsburgh campaign associated with Madhani et al. [1], where a solar-filtered balloon camera recorded partial eclipse phases while the payload moved.

The gradient-based solar edge analysis provides a deliberately simple quantitative comparison without requiring absolute photometry. Defining the apparent solar edge from the strongest two-dimensional intensity transition is less dependent on absolute image brightness than a fixed intensity threshold and is therefore attractive for a sequence in which source brightness and imaging conditions vary. The strong correlation with the independently calculated eclipse geometry indicates that quantitative information on the changing solar crescent can be recovered even from these small, non-stabilized wide-angle images. The remaining differences are, however, substantial enough that the present method should be interpreted as a proof of concept rather than as calibrated solar-limb metrology.

Several effects can shift or broaden the measured maximum-gradient edge region. The apparent solar disk occupies only a small number of pixels, so the result is sensitive to the camera point-spread function, residual motion blur, JPEG processing and sampling. The two halves of the sequence were obtained with two nominally identical cameras, but small camera-to-camera differences in focus, lens response or filter placement cannot be excluded. In addition, the solar image moved through different parts of a wide-angle lens, for which vignetting and field-dependent aberrations are expected. The eclipse-glass filter was mounted in front of an action-camera optical path rather than in a calibrated astronomical filter holder; changes in incidence angle may therefore modify both transmission and the effective image profile. At the low solar elevations of the later observations, atmospheric scattering may further broaden the apparent image. Also changing weather conditions such as high-altitude clouds may influence the results. These effects were deliberately not corrected because the available data do not permit them to be separated reliably; instead, they define the instrumental calibration programme for the planned follow-up experiments.

The normalization to the first frame is another important limitation. It removes the need to know the unocculted solar image area in pixels, but it transfers any bias in the first measurements directly to all subsequent values. Likewise, the analysis assumes that changes in the measured connected area primarily reflect lunar occultation. The irregular deviations between neighbouring frames show that this assumption is only approximate. Nevertheless, using a single predefined image-processing approach for the complete sequence avoids frame-by-frame tuning to the expected eclipse geometry.

The temporal sampling is intrinsically stochastic. Although images were acquired every 10 s, the Sun was recorded only when the instantaneous payload attitude placed it within the filtered camera field. Consequently, useful frames are separated by intervals ranging from seconds to many minutes, and no image was obtained exactly at the local maximum. This is the principal trade-off of using passive payload rotation instead of an active solar-pointing system. The balloon itself later reached approximately 35 km altitude, but the solar sequence begins immediately before release and continues during ascent. It should therefore be described as balloon-borne eclipse imaging during a high-altitude balloon flight rather than as a sequence acquired entirely from the stratosphere.

A dedicated follow-up study is planned to convert this proof-of-concept approach into a calibrated measurement method. Ground-based measurements with the same camera-filter combinations before flight could quantify camera-specific lens vignetting, field-dependent point-spread functions, geometric distortion, and filter transmission as a function of incidence angle. Simultaneous solar imaging with both cameras would separate camera-to-camera response from temporal changes in atmospheric conditions, while an unocculted reference Sun would provide an independent absolute-area calibration. RAW rather than JPEG acquisition, a larger solar image, and inertial attitude data would further reduce processing and sampling uncertainties. These calibrations would make it possible to test whether the gradient-derived edge can recover geometrical obscuration with substantially smaller systematic error while retaining the mechanically simple passive-rotation concept.

More generally, low-cost balloon observations occupy an interesting position between professional eclipse expeditions and ground-based citizen-science observations. The hardware used here was commercially available, the observing concept required no dedicated mechanical stabilization, and the recovered data can be interpreted directly by students and the public. The Spanish 2026 balloon missions provide a valuable complementary dataset, with early public results emphasizing wide-field views of totality, the atmospheric horizon and the lunar shadow [4]. More detailed comparison will become possible as technical reports and image sequences from those flights are published.

## 5. Conclusions

A solar-filtered Insta360 ONE RS wide-angle imaging system on a freely rotating high-altitude balloon recorded 21 direct images of the partial solar eclipse of 12 August 2026. The sequence spans both sides of eclipse maximum despite the absence of active pointing. A simple two-dimensional gradient-based solar edge analysis recovered the expected eclipse progression with a Pearson correlation of 0.959 relative to independent geometrical calculations. The present study is therefore best regarded as a proof of concept showing that quantitative eclipse information can be extracted from mechanically simple, non-stabilized balloon imagery. The residual systematic and frame-to-frame differences identify the principal targets for the next experimental step: camera-specific calibration of point-spread function, lens vignetting and geometric distortion, filter-angle response, and simultaneous cross-calibration of the two cameras. With these additions, the passive wide-field concept could develop into a quantitatively calibrated method for future balloon-borne eclipse observations.

#### Acknowledgements

The authors gratefully acknowledge Freie Schule Oldenburg for providing access to the school grounds and supporting the launch of the high-altitude balloon. The location proved particularly suitable for the experiment, as it provided sufficient space for launch preparations while allowing the balloon to leave the immediate urban environment of Oldenburg shortly after release.

We also gratefully acknowledge LEB Niedersachsen for supporting the Tiny Observatory project. The technical infrastructure developed within Tiny Observatory provided an important basis for the preparation and implementation of the balloon experiments.

#### Declaration of AI-assisted work

ChatGPT (OpenAI) was used to assist with language editing and the development and testing of Python-based image-analysis procedures.